\documentclass{article}%
\usepackage{amsmath}
\usepackage{amsfonts}
\usepackage{amssymb}
\usepackage{graphicx}%
\providecommand{\U}[1]{\protect\rule{.1in}{.1in}}

\begin{document}

\begin{center}
\textbf{THE VACUOLE MODEL: NEW TERMS IN THE SECOND ORDER DEFLECTION OF LIGHT }

\bigskip

Amrita Bhattacharya$^{1,a}$, Guzel M. Garipova$^{2,b}$, Ettore Laserra$^{3,c}$,

Arunava Bhadra$^{4,d}$ and Kamal K. Nandi$^{1,2,e}$

$\bigskip$

$^{1}$Department of Mathematics, University of North Bengal, Siliguri 734 013, India

$^{2}$Department of Theoretical Physics, Sterlitamak State Pedagogical
Academy, Sterlitamak 453103, Russia

$^{3}$DMI,\ Universit\`{a} di Salerno, Via Ponte Don Melillo, 84084 Fisciano,
Salerno, Italy

$^{4}$High Energy and Cosmic Ray Research Center, University of North Bengal,
Siliguri 734 013, India

$\bigskip$

$\bigskip$

$^{a}$Email: amrita\_852003@yahoo.co.in

$^{b}$Email: goldberg144@gmail.com

$^{c}$Email: elaserra@unisa.it

$^{d}$Email: aru\_bhadra@yahoo.com

$^{e}$E-mail: kamalnandi1952@yahoo.co.in

\bigskip
\end{center}


\begin{center}
-----------------------------------------------------------------------------------

\bigskip

\textbf{Abstract}
\end{center}

The present paper is an extension of a recent work (Bhattacharya et al. 2010)
to the Einstein-Strauss vacuole model with a cosmological constant, where we
work out the light deflection by considering perturbations up to order $M^{3}$
and confirm the light bending obtained previously in their vacuole model by
Ishak et al. (2008). We also obtain another local coupling term $-\frac{5\pi
M^{2}\Lambda}{8}$ related to $\Lambda$, in addition to the one obtained by
Sereno (2008, 2009). We argue that the vacuole method for light deflection is
exclusively suited to cases where the cosmological constant $\Lambda$
disappears from the path equation. However, the original Rindler-Ishak method
(2007) still applies even if a certain parameter $\gamma$ of Weyl gravity does
not disappear. Here, using an alternative prescription, we obtain the known
term $-\frac{\gamma R}{2}$, as well as another new local term $\frac
{3\pi\gamma M}{2}$ between $M$ and $\gamma$. Physical implications are
compared, where we argue that the repulsive term $-\frac{\gamma R}{2}$ can be
\textit{masked} by the Schwarzschild term $\frac{2M}{R}$ in the halo regime
supporting attractive property of the dark matter.

\begin{center}
---------------------------------------------------------------------------------------
\end{center}

\textbf{I. Introduction}

Recently, we confirmed the Rindler-Ishak method (Rindler \& Ishak, 2007) by
calculating light bending in a more general solution, viz., the
Mannheim-Kazanas-de Sitter solution of Weyl conformal gravity (Bhattacharya et
al. 2010) that contains two parameters $\Lambda$ and $\gamma$, the latter is
assumed to play a prominent role in the galactic halo populated by dark
matter. The method indeed delivered the effect of $\gamma$ exactly as it has
been known in the literature for long. Subsequently, the $\Lambda-$ effect has
been calculated by Ishak et al. (2008) within the framework of
Einstein-Strauss vacuole (its earlier incarnation is the Kottler vacuole). Our
broad aim here is to examine how the effects of both $\Lambda$ and $\gamma$
appear from suitable considerations in the Weyl conformal gravity.

Galaxies or clusters of galaxies (hereinafter called lenses for brevity) have
been modelled as residing in the womb of a de Sitter vacuole much larger than
their sizes (Ishak et al. 2008). The model assumes that, for a given lens, the
boundary radius $r_{b}$ of the vacuole is determined where the spacetime
transitions from a Schwarzschild-de Sitter (SdS) spacetime to a cosmological
Friedman-Robertson-Walker (FRW) background. Further, all the light-bending
occurs in the SdS vacuole and that once the light transitions out of the
vacuole and into FRW spacetime, all bending stops. Ishak et al. (2008) showed
that the effect of the cosmological constant $\Lambda$ appears inside the
vacuole in the bending of light by different lens systems. They also obtained
an upper bound on $\Lambda$, using observational uncertainties in the
measurement of the bending of light, which turned out to be only two orders of
magnitude away from the cosmologically determined value. For a lens of mass
$M$ and radius $R$, Ishak et al. (2008) obtained light deflection up to second
order in $\frac{M}{R}$ together with a $\Lambda-$repulsion term ($=-\Lambda
Rr_{b}/6$).

The purpose of the present article is to confirm the light deflection in the
second order by using perturbations up to third order\footnote{We are indebted
to an anonymous referee for suggesting that the correct deflection follows
from (at least) third order calculation.}. When we do that we come up with new
extra terms, while confirming the most interesting term $-\Lambda Rr_{b}/6$.
Additionally, one might like to have an idea of how the presence of a
conformal parameter $\gamma$ would affect light deflection by the lenses. To
this end, we use an alternatuive orescription for the more general exact
Mannheim-Kazanas-de Sitter solution of Weyl conformal gravity that includes
the parameter $\gamma$. Pure SdS vacuole with only $M$ and $\Lambda$ is
readily recovered at $\gamma=0$. New coupling terms arising out of the
invariant angle have been obtained.

\textbf{II. The solution and the approximation scheme}

One well discussed solution that contains both the conformal $\gamma$ and dS
$\Lambda$ effects is the Mannheim-Kazanas-de Sitter (MKdS) solution (Mannheim
\& Kazanas 1989; Mannheim 1997, 2006) of Weyl conformal gravity field
equations. The metric is given, in units $G=c_{0}=1$, by (see e.g., Edery \&
Paranjape 1998):%
\begin{align}
d\tau^{2}  &  =B(r)dt^{2}-A(r)dr^{2}-r^{2}(d\theta^{2}+\sin^{2}\theta
d\varphi^{2}),\text{ \ }\\
B(r)  &  =A^{-1}(r)=1-\frac{2M}{r}+\gamma r-\frac{\Lambda}{3}r^{2},
\end{align}
where $M$ is the central mass, $\Lambda$ and $\gamma$ are constants. The
accepted numerical value from current cosmology is $\Lambda=1.29\times
10^{-56}$cm$^{-2}$. However, there seems to be some ambiguity about the sign
and magnitude of $\gamma$. Mannheim and Kazanas fix it from flat rotation
curve data to be positive and of the order of the inverse Hubble length, while
Pireaux (2004) argues for $\left\vert \gamma\right\vert \sim10^{-33}$
cm$^{-1}$. Edery \& Paranjape (1998) obtained a negative value from the
gravitational time delay by galactic clusters. We shall keep the value of
$\gamma$ open, but for purely illustrative purposes, take the value
$\gamma=3.06\times10^{-30}$ cm$^{-1}$ (Mannheim 2006). We emphasize that the
methods adopted here do not need to assume any particular value of $\gamma-$
it is essentially kept free to be fixed by more accurate observations.

In the null geodesic equation, $\Lambda$ cancels out giving%
\begin{equation}
\frac{d^{2}u}{d\varphi^{2}}=-u+3Mu^{2}-\frac{\gamma}{2}.
\end{equation}
We have recently solved the light ray equation using the Rindler-Ishak method
(Rindler \& Ishak 2007), and have shown how $\gamma$ and $\Lambda$ get mixed
up in the deflection at higher order (Bhattacharya et al. 2010). Usual
perturbative expansion up to order $M^{3}$ gives the final solution of Eq.(3)
as:%
\begin{align}
u  &  \equiv\frac{1}{r}=\frac{\sin\varphi}{R}-\frac{\gamma}{2}\nonumber\\
&  +\frac{M}{4R^{2}}\left[  6+3R^{2}\gamma^{2}-3R\gamma(\pi-2\varphi
)\cos\varphi+2\cos2\varphi-6R\gamma\sin\varphi\right] \nonumber\\
&  -\frac{3M^{2}}{32R^{3}}[96R\gamma+24R^{3}\gamma^{3}-10(2+3R^{2}\gamma
^{2})(\pi-2\varphi)\cos\varphi+32R\gamma\cos2\varphi\nonumber\\
&  -20\sin\varphi-30R^{2}\gamma^{2}\sin\varphi+3\pi^{2}R^{2}\gamma^{2}%
\sin\varphi-12\pi R^{2}\gamma^{2}\varphi\sin\varphi\nonumber\\
&  +12R^{2}\gamma^{2}\varphi^{2}\sin\varphi-8\pi R\gamma\sin2\varphi
+16R\gamma\varphi\sin2\varphi+2\sin3\varphi]\nonumber\\
&  +\frac{M^{3}}{128R^{4}}[9R\gamma(\pi-2\varphi)\{R^{2}\gamma^{2}(\pi
^{2}-150-4\pi\varphi+4\varphi^{2})-260\}\cos\varphi\nonumber\\
&  +2\{816+2916R^{2}\gamma^{2}-36R^{2}\gamma^{2}(\pi^{2}-27-4\pi
\varphi+4\varphi^{2})\cos2\varphi\nonumber\\
&  +27R\gamma(\pi-2\varphi)\cos3\varphi-4\cos4\varphi-1170R\gamma\sin
\varphi+540R^{4}\gamma^{4}+356\cos2\varphi\nonumber\\
&  +108\pi^{2}R^{3}\gamma^{3}\sin\varphi-360\pi R\gamma\varphi\sin
\varphi-432\pi R^{3}\gamma^{3}\varphi\sin\varphi+360R\gamma\varphi^{2}%
\sin\varphi\nonumber\\
&  +432R^{3}\gamma^{3}\varphi^{2}\sin\varphi-120\pi\sin2\varphi-396\pi
R^{2}\gamma^{2}\sin2\varphi+90\pi^{2}R\gamma\sin\varphi\nonumber\\
&  +792R^{2}\gamma^{2}\varphi\sin2\varphi-675R^{3}\gamma^{3}\sin
\varphi+240\varphi\sin2\varphi+126R\gamma\sin3\varphi\}],
\end{align}
where $R$ is related to the closest distance approach $r_{0}$ defined by
($\varphi=\pi/2$ $)$%
\begin{align}
\frac{1}{r_{0}}  &  =\frac{1}{16R^{3}}[4MR(4-6R\gamma+3R^{2}\gamma^{2}%
)-8R^{2}(R\gamma-2)\nonumber\\
&  +M^{2}(33-96R\gamma+45R^{2}\gamma^{2}-36R^{3}\gamma^{3})]\nonumber\\
&  +\frac{3M^{3}}{64R^{4}}[152-432R\gamma+648R^{2}\gamma^{2}-225R^{3}%
\gamma^{3}+180R^{4}\gamma^{4}].
\end{align}
Note that Eq.(4) is a more general solution involving $\gamma$ and we can
recover the SdS vacuole case putting $\gamma=0$. The Rindler-Ishak method
requires another function $A(r,\varphi)\equiv\frac{dr}{d\varphi}$, which
yields for the present solution
\begin{align}
A(r,\varphi)  &  =(-r^{2})\times\nonumber\\
&  [\frac{\cos\varphi}{32R^{3}}(32R^{2}-3M^{2}\{20+3R^{2}\gamma^{2}%
(10+(\pi-2\varphi)^{2}\}\nonumber\\
&  +32MR(9M\gamma-2)\sin\varphi)-6M\{3M\cos3\varphi-8MR\gamma(\pi
-2\varphi)\cos2\varphi\nonumber\\
&  +(10M-4R^{2}\gamma+9MR^{2}\gamma^{2})(\pi-2\varphi)\sin\varphi
\}]\nonumber\\
&  +M^{3}[18R\gamma\{130+\pi^{2}(10+9R^{2}\gamma^{2})-4\pi\varphi
(10+9R^{2}\gamma^{2})+40\varphi^{2}\nonumber\\
&  +3R^{2}\gamma^{2}(25+12\varphi^{2})\}\cos\varphi-48(10+27R^{2}\gamma
^{2})(\pi-2\varphi)\cos2\varphi\nonumber\\
&  +648R\gamma\cos3\varphi+1620R\gamma\pi\sin\varphi+486\pi R^{3}\gamma
^{3}\sin\varphi-9\pi^{3}R^{3}\gamma^{3}\sin\varphi\nonumber\\
&  +3240R\gamma\varphi\sin\varphi-972R^{3}\gamma^{3}\varphi\sin\varphi
+54\pi^{2}R^{3}\gamma^{3}\varphi\sin\varphi\nonumber\\
&  -108\pi R^{3}\gamma^{3}\varphi^{2}\sin\varphi+72R^{3}\gamma^{3}\varphi
^{3}\sin\varphi-944\sin2\varphi-2304R^{2}\gamma^{2}\sin2\varphi\nonumber\\
&  +144\pi^{2}R^{2}\gamma^{2}\sin2\varphi-576\pi R^{2}\gamma^{2}\varphi
\sin2\varphi+576R^{2}\gamma^{2}\varphi^{2}\sin2\varphi\nonumber\\
&  -162\pi R\gamma\sin3\varphi+324R\gamma\varphi\sin3\varphi+32\sin4\varphi].
\end{align}
Assume a small angle $\varphi_{b}$ at the vacuole boundary radius $r=r_{b}$
such that $\sin\varphi_{b}\simeq\varphi_{b}$, and $\cos\varphi_{b}\simeq1$.
Then the above gives
\begin{align}
\frac{1}{r_{b}}  &  =-\frac{\gamma}{2}+\frac{\varphi_{b}}{R}+M\left[  \frac
{2}{R^{2}}-\frac{3\pi\gamma}{4R}+\frac{3\gamma^{2}}{4}\right] \nonumber\\
&  +M^{2}\left[  \frac{15\pi}{8R^{3}}-\frac{12\gamma}{R^{2}}+\frac{45\pi
\gamma^{2}}{16R}-\frac{9\gamma^{3}}{4}-\frac{39\varphi_{b}}{16R^{3}}%
+\frac{3\pi\gamma\varphi_{b}}{2R^{2}}\right. \nonumber\\
&  \left.  -\frac{45\gamma^{2}\varphi_{b}}{16R}-\frac{9\pi^{2}\gamma
^{2}\varphi_{b}}{32R}-\frac{3\gamma\varphi_{b}^{2}}{R^{2}}+\frac{9\pi
\gamma^{2}\varphi_{b}^{2}}{8R}-\frac{9\gamma^{2}\varphi_{b}^{3}}{8R}\right]
\nonumber\\
&  +M^{3}\left[  \frac{73}{4R^{4}}-\frac{1143\pi\gamma}{64R^{3}}%
+\frac{243\gamma^{2}}{4R^{2}}-\frac{9\pi^{2}\gamma^{2}}{16R^{2}}-\frac
{675\pi\gamma^{3}}{64R}\right. \nonumber\\
&  \left.  +\frac{9\pi^{3}\gamma^{3}}{128R}+\frac{135\gamma^{4}}{16}%
-\frac{15\pi\varphi_{b}}{4R^{4}}+\frac{747\gamma\varphi_{b}}{32R^{3}}%
+\frac{45\pi^{2}\gamma\varphi_{b}}{32R^{3}}\right. \nonumber\\
&  \left.  +\frac{15\varphi_{b}^{2}}{2R^{4}}-\frac{81\pi\gamma^{2}\varphi_{b}%
}{8R^{2}}+\frac{675\gamma^{3}\varphi_{b}}{64R}+\frac{81\pi^{2}\gamma
^{3}\varphi_{b}}{64R}-\frac{45\pi\gamma\varphi_{b}^{2}}{8R^{3}}\right.
\nonumber\\
&  \left.  +\frac{45\gamma^{2}\varphi_{b}^{2}}{2R^{2}}-\frac{189\pi\gamma
^{3}\varphi_{b}^{2}}{32R}+\frac{45\gamma\varphi_{b}^{3}}{8R^{3}}%
+\frac{99\gamma^{3}\varphi_{b}^{3}}{16R}\right]
\end{align}
and Eq.(6) gives%
\begin{align}
A_{b}  &  \equiv A(r_{b},\varphi_{b})=\frac{r_{b}^{2}}{R}-\frac{2M\varphi
_{b}r_{b}^{2}}{R^{2}}+\frac{M^{2}r_{b}^{2}}{32R^{3}}[48\pi R\gamma
-78-90R^{2}\gamma^{2}\nonumber\\
-9\pi^{2}R^{2}\gamma^{2}  &  -192R\gamma\varphi_{b}+72\pi R^{2}\gamma
^{2}\varphi_{b}-108R^{2}\gamma^{2}\varphi_{b}^{2}]+O(M^{3}),
\end{align}
where the terms $O(M^{3})$ are straightforward but rather lengthy and hence
not displayed here.

Note that observations give us values of $M$ and $R$ for a lens, but we have
only one equation (7) for two unknowns $\varphi_{b}$ and $r_{b}$. Hence we
need to specify any one of them from independent considerations. Along with
Ishak et al. (2008), we shall employ Einstein-Strauss prescription (Einstein
\& Strauss 1945; Schucking 1954) to determine $r_{b}$ assuming that the
vacuole has been matched to an expanding FRW universe via the
Sen-Lanczos-Darmois-Israel junction conditions (Sen 1924; Lanczos 1924;
Darmois 1927; Israel 1966). In general, the vacuole radius $r_{b}$ would also
change due to cosmic expansion, but we shall consider $r_{b}$ at that
particular instant $t_{0}$ of cosmic epoch when the light ray happens to pass
the point of closest approach to the lens. Thus $r_{b}$ is determined by
exploiting the Einstein-Strauss prescription [see Ishak et al. (2008)]
\begin{equation}
r_{b\text{ in SdS}}=a(t)r_{b\text{ in FRW}}\text{, }M_{\text{SdS}}=\frac{4\pi
}{3}r_{b\text{ in SdS }}^{3}\times\rho_{\text{in FRW}}.
\end{equation}
To achieve exact matching with the exterior FRW universe, the energy density
$\rho$ within the vacuole should have a contribution from $\Lambda$ besides
that of ordinary matter $\rho_{\text{m}}$, that is, $\rho=\rho_{\text{m}}%
+\rho_{\Lambda}=$ $\frac{3H_{0}^{2}}{8\pi}(\Omega_{\text{m}}+\Omega_{\Lambda
})$, where $\Omega_{\text{m}}=8\pi\rho_{\text{m}}/3H_{0}^{2}$, $\Omega
_{\Lambda}=8\pi\rho_{\Lambda}/3H_{0}^{2}$ are the matter and dark energy
densities in dimensionless form. Current observations suggest that the
universe is spatially flat so that $\rho=\rho_{\text{critical }}=\frac
{3H_{0}^{2}}{8\pi}$, which in turn imply that $\Omega=\Omega_{\text{m}}%
+\Omega_{\Lambda}=1$. Type Ia supernova observations yield $\Omega_{\text{m}%
}=0.27$ so that $\Omega_{\Lambda}=0.73$ (Riess et al. 1998; Perlmutter et al.
1999; Carroll 2001; Page et al. 2003; Peebles \& Ratra 2003; Spergel et al.
2007). For computational purposes, we shall take the density to be
$\rho_{\text{in FRW}}=\rho_{\text{critical}}=\frac{3H_{0}^{2}}{8\pi}%
=1.1\times10^{-29}\left(  \frac{H_{0}}{75\text{ km/sec/Mpc}}\right)  ^{2}%
$gm/cm$^{3}$ (Weinberg 1972) inside the vacuole. A slight deviation from this
value would not drastically alter our conclusions. Normalizing the scale
factor to $a(t_{0})=1$ and dropping suffixes, the above prescription
translates to
\begin{equation}
r_{b}=\left(  \frac{3M}{4\pi\rho}\right)  ^{1/3},
\end{equation}
where $M$ is the lens mass often expressed in units of sun's mass $M_{\odot
}=1.989\times10^{33}$ gm $=1.475\times10^{5}$cm.

We should now solve the cubic Eq.(7) in $\varphi_{b}$ to find three roots
designated by $\varphi_{b}^{i}=\varphi_{b}^{i}(\rho,M,R,\gamma)$ where
$i=1,2,3$. For the SdS vacuole ($\gamma=0$), we fortunately get only a single
root $\varphi_{b}$, which becomes, using the Einstein-Strauss prescription for
$r_{b}$ from Eq.(10),
\begin{equation}
\varphi_{b}=\frac{90\pi M^{3}+96RM^{2}-16(\pi\rho)^{1/3}(6M)^{2/3}R^{3}%
}{117M^{3}-48MR^{2}}.
\end{equation}
Our interest is to express the deflection angle $\psi$ in terms of $M,R$ and
an as yet unspecified $\rho$. We shall use the above value of $\varphi_{b}$
later. The Rindler-Ishak formula for $\psi$ at $r=r_{b}$ is%
\begin{equation}
\tan\psi=\frac{r_{b}\sqrt{B(r_{b})}}{\left\vert A_{b}\right\vert }.
\end{equation}
From Eq.(7), one sees that $r_{b}$ contains the radius $R$ which is a real
root of Eq.(5) \ For large distances, there is little difference between $R$
and $r_{0}$. So we shall identify $R$ with the Einstein radius where the
closest approach distance $r_{0}$ appears.\footnote{The half angle of the
Einstein ring subtended at the observer is defined as $\theta_{E}=2\epsilon
D_{ls}/D_{os},$ the suffixes $o,l,s$ in $D$ representing angular diameter
distances between observer, lens and the source, assuming all to be situated
on a "line". We have taken $R=R_{E}=D_{ol}\tan\theta_{E}\simeq D_{ol}%
\theta_{E}$ for small $\theta_{E}$.} We are not using the impact paramter here.

Our general algorithm for calculation proceeds along the following analytical steps:

(1) Put the expression for $r_{b}$ from Eq.(7) and $A_{b}$ from Eq.(8) into
Eq.(12) for deflection angle.

(2) Expand the right hand side of Eq.(12) in first power of $\gamma$ in order
to separate out its contribution from the SdS one. Thus we write formally%
\begin{equation}
\tan\psi^{\text{total}}\simeq C(\varphi_{b},M,R)+\gamma D(\varphi_{b},M,R)
\end{equation}
where $C$, $D$ are known functions to be expanded in powers of $M$. For small
$\psi^{\text{total}}$, we decompose%
\begin{equation}
\tan\psi^{\text{total}}\simeq\psi^{\text{total}}=\psi^{\text{SdS}}%
+\psi^{\text{MKdS}}.
\end{equation}

(3) Expand both $C$ and $D$ up to the power $M^{2}$ to see the individual
contributions of terms.

\textbf{III. SdS vacuole: }$\Lambda-$\textbf{effect}

This case corresponds to $\gamma=0$ and we have to be concerned with only
$C=\tan\psi^{\text{SdS}}\simeq\psi^{\text{SdS}}$. The boundary radius $r_{b}$
of the vacuole from Eq.(7) is [step (1)],
\begin{equation}
r_{b}=\frac{16R^{4}}{32MR^{2}+16R^{3}\varphi_{b}+M^{2}(30\pi R-39R\varphi
_{b})+M^{3}(292-60\pi\varphi_{b}+120\varphi_{b}^{2})}%
\end{equation}
and Eq.(8) gives%
\begin{align}
A_{b}  &  =\frac{r_{b}^{2}}{R}-\frac{2Mr_{b}^{2}\varphi_{b}}{R^{2}}%
-\frac{M^{2}r_{b}^{2}}{R^{3}}\left\{  \frac{39}{16}+\frac{15\pi\varphi_{b}}%
{8}-\frac{15\varphi_{b}^{2}}{4}\right\} \nonumber\\
&  -\frac{M^{3}r_{b}^{2}}{R^{4}}\left\{  \frac{15\pi}{4}+\frac{25\varphi_{b}%
}{4}\right\}  .
\end{align}
Note that Eqs.(10), (11) and (15) are consistent. Putting the values of
$r_{b}$ and $A_{b}$ in Eq.(12), assuming a small $\psi^{\text{SdS}}$, and
expanding in powers of $M$ we get [step (2)]%
\begin{align}
\psi^{\text{SdS}}  &  =\left(  1-\frac{\Lambda R^{2}}{6\varphi_{b}^{2}%
}\right)  \varphi_{b}+M\left[  \varphi_{b}\left(  \frac{2R\Lambda}%
{3\varphi_{b}^{2}}-\frac{\varphi_{b}}{R}\right)  +\left(  1-\frac{\Lambda
R^{2}}{6\varphi_{b}^{2}}\right)  \left(  \frac{2}{R}+\frac{2\varphi_{b}^{2}%
}{R}\right)  \right] \nonumber\\
&  +M^{2}\left[  \left(  \frac{5\pi\Lambda}{8\varphi_{b}^{3}}-\frac{13\Lambda
}{16\varphi_{b}^{2}}-\frac{2}{R^{2}}-\frac{2\Lambda}{\varphi_{b}^{4}}\right)
\varphi_{b}+\left(  \frac{2R\Lambda}{3\varphi_{b}^{3}}-\frac{\varphi_{b}}%
{R}\right)  \left(  \frac{2}{R}+\frac{2\varphi_{b}^{2}}{R}\right)  \right.
\nonumber\\
&  \left.  -\left(  1-\frac{\Lambda R^{2}}{6\varphi_{b}^{2}}\right)  \left\{
-\frac{4\varphi_{b}}{R^{2}}-\frac{240\pi R-312R\varphi_{b}}{128R^{3}}\right.
\right. \nonumber\\
&  \left.  \left.  +128R^{3}\varphi_{b}\left(  -\frac{39}{2048R^{5}}%
-\frac{15\pi\varphi_{b}}{1024R^{5}}-\frac{\varphi_{b}^{2}}{512R^{5}}\right)
\right\}  \right]  +O(M^{3}).
\end{align}
To simplify calculations, we shall now expand $\varphi_{b}$ of Eq.(11) with
$\rho=\frac{3M}{4\pi r_{b}^{3}}$ obtaining%
\begin{equation}
\varphi_{b}=\frac{R}{r_{b}}-\frac{2M}{R}+M^{2}\left(  \frac{39}{16Rr_{b}%
}-\frac{15\pi}{8R^{2}}\right)  .
\end{equation}
Using it in $\psi^{\text{SdS}}$, and collecting terms of similiar orders in
$M$, we get%
\begin{align}
\psi^{\text{SdS}}  &  \simeq\varphi_{b}-\frac{\Lambda Rr_{b}}{6}+\frac{M}%
{R}\left[  2+\frac{R^{2}}{r_{b}^{2}}-\frac{R^{2}\Lambda}{3}\right] \nonumber\\
&  +\frac{M^{2}}{R^{2}}\left[  \frac{15\pi}{8}-\frac{7R^{3}}{4r_{b}^{3}}%
+\frac{15\pi R^{2}}{8r_{b}^{2}}-\frac{4R}{r_{b}}-\frac{5\pi R^{2}\Lambda}%
{16}-\frac{R^{3}\Lambda}{24r_{b}}+\frac{25\Lambda Rr_{b}}{96}\right]
\end{align}
As usual, for small angle, $\tan\varphi_{b}\simeq\varphi_{b}$, so that the
deflection $\epsilon^{\text{SdS}}$ for nonzero $\varphi_{b}$ is, by definition
(Rindler \&\ Ishak 2007)%
\begin{equation}
\epsilon^{\text{SdS}}=\tan(\psi^{\text{SdS}}-\varphi_{b})\simeq\psi
^{\text{SdS}}-\varphi_{b}.
\end{equation}
Assuming that $r_{b}>>R$, $R>>M$, and collecting terms of interest, we get the
total deflection
\begin{equation}
2\epsilon^{\text{SdS}}=-\frac{\Lambda Rr_{b}}{3}+\frac{4M}{R}+\frac{15\pi
M^{2}}{4R^{2}}-\frac{2M\Lambda R}{3}-\frac{5\pi M^{2}\Lambda}{8}+\frac
{2MR}{r_{b}^{2}}.
\end{equation}
Clearly, the above yields the Ishak et al term $-\frac{\Lambda Rr_{b}}{6}$ as
well as the well known Schwarzschild terms. Interestingly, identifying the
constant $R\simeq r_{0}$ from Eq.(5), the fourth term, viz., $t_{4}%
^{\text{l.c.}}=-\frac{2M\Lambda r_{0}}{3}$ looks numerically like the same as
what Sereno (2009) calls the \textit{local coupling term }between $M$ and
$\Lambda$\textit{, }since the term does not depend on the vacuole radius
$r_{b}$ (see also the Appendix). Nonetheless, it does depend on the particular
path via $R$. Remarkably, we also discover a \textit{new} local coupling term
in the second order, viz., $t_{5}^{\text{l.c.}}=-\frac{5\pi M^{2}\Lambda}{8}$,
coming from the fifth term in the second square bracket in Eq.(19). This seems
to be a more genuine local coupling term because it does not involve the
parameter $R$ of the light trajectory. However, both the terms contribute
repulsively to bending.

A certain thing is to be noted here. We might start with the first order
differential equation already containing $\Lambda$ through $B(r)$:%
\begin{equation}
\frac{1}{r^{4}}\left(  \frac{dr}{d\varphi}\right)  ^{2}+\frac{B(r)}{r^{2}%
}-\frac{1}{b^{2}}=0,
\end{equation}
where $b$ is the impact parameter defined as $\ell/e$. Then one can define $b$
using the closest approach distance $r=r_{0}$, where $\frac{dr}{d\varphi}=0$,
which yields from the first order Eq.(22) the value of $b$ as%
\begin{equation}
b=r_{0}\left[  \frac{1}{B(r_{0})}\right]  ^{1/2}.
\end{equation}
This gives%
\begin{align}
r_{0}  &  \simeq b\left(  1-\frac{M}{b}-\frac{b^{2}\Lambda}{6}\right)  .\\
\frac{1}{r_{0}}  &  \simeq\frac{1}{b}\left(  1+\frac{M}{b}+\frac{b^{2}\Lambda
}{6}+\frac{M\Lambda b}{3}\right)
\end{align}
Using the values of $R=r_{0}$ and $1/R=1/r_{0}$ from above into the expression
(), we find the relevant terms to add to%
\begin{equation}
2\epsilon^{\text{SdS}}=-\frac{b\Lambda r_{b}}{3}+\frac{4M}{b}+\frac{2M\Lambda
b}{3}-\frac{2M\Lambda b}{3}+\text{terms in }M^{2},
\end{equation}
so that the local coupling term $\frac{2M\Lambda b}{3}$ cancels out!

In our opinion, some caution should be exercised in the interpretation that
such a local coupling really vanishes due to that cancellation. Note that both
the original Rindler-Ishak (2007) or vacuole (2008) method do not at all use
the first order path equation in which $\Lambda$ already appears. Their whole
package consists of the second order differential equation in which $\Lambda$
does not appear (which led people to believe that it does not hence affect
light bending) and the definition of the invariant angle $\psi$ to capture the
effect of $\Lambda$, without needing any further ingredients. To use the first
order path equation (22) \textit{already }containing $\Lambda$ at any stage of
the present calculation would mean capturing the effect of $\Lambda$ twice. We
argue that both the trajectory equation (3) with $R\simeq r_{0}$ and the first
order equation (22) with $b$ should not be simultaneously used. Integration of
the first order equation (22) with $b$ and the present vacuole method should
be treated as mutually exclusive ways, both separately yielding the bending
with expected local coupling terms. We obtained not only the Sereno-like local
coupling term $-\frac{2M\Lambda r_{0}}{3}$ but also a new term $-\frac{5\pi
M^{2}\Lambda}{8}$ including other new terms, the most notable one being
$-\frac{\Lambda Rr_{b}}{3}$.

As argued above, the vacuole method has been exclusively tailored to capture
the effect of a parameter (like $\Lambda$) that has disappeared from the path
equation. To justify this statement, we might try to assess the influence of
the remaining parameter $\gamma$, which has \textit{not }disappeared from the
second order equation (3). If we still apply the present vacuole method, it
can be verified that the known effect $-\frac{\gamma R}{2}$ on bending does
not appear in $\psi^{\text{MKdS}}$ (We omit the details). However, we can
still use the initial Rindler-Ishak non-vacuole method (2007) to obtain
exactly this effect as already shown in Bhattacharya et al. (2010). Below we
shall show that even a reverse prescription in the non-vacuole method yields
the same effect $-\frac{\gamma R}{2}$.

\textbf{V. The }$\gamma$\textbf{ effect by alternative prescription}

To capture the effect of $\gamma$ by means of the invariant angle, we assume
here for simplicity $\Lambda=0$ and restrict to first order in $M$. In
principle, this case corresponds to $C=0$ and we have to be concerned with
only
\begin{equation}
\gamma D=\tan\psi^{\text{MKdS}}\simeq\psi^{\text{MKdS}}.
\end{equation}
But as stated above, $\gamma D$ does not yield the effect $-\frac{\gamma R}%
{2}$. Therefore we proceed as follows.

First, a few words about the original Rindler-Ishak (2007) prescription and
the alternative one we are going to follow here. In the asymptotically
non-flat metric, the limit $r\rightarrow\infty$ makes no sense. Therefore they
prescribed that the only intrinsically characterized $r$ value replacing
$r\rightarrow\infty$ is the one at $\varphi=0$. The measurable quantities are
the various $\psi$ angles that the photon orbit makes with successive
coordinate planes $\varphi=$ const. While this assumption is perfectly valid
giving the desired results, we shall implement the Rindler-Ishak method
following a \textit{reverse} prescription\footnote{There seems to be no
obvious operative way of actually measuring the azimuthal angle on the sky
unless one is able to perform the experiment twice, once with the lens in the
way, and once without (just as is done with gravitational bending of light by
the sun). However, one can define a distance $r$ on the orbit by using a
coordinate system in which the lens is at $r=0$. Then for an observer far
enough beyond the lens, the very fact that light reaches the observer entails
that the point $r=\infty$ does lie on the geodesic. The form of the geodesic
then determines at what value of $\varphi\neq0$ the source lies. This is what
we are calling reverse prescription here.}, namely, determining $\varphi\neq0$
occurring at a point $r=\infty$ on the null orbit and work out how the
parameters $\gamma$ and $M$ appear in the light bending.

Next, for the unbound orbits associated with lensing, the distance of closest
approach of a light ray to a galaxy will be further from the center of the
galaxy than the matter orbiting inside it. Hence our goal here is to calculate
the deflection angle $\epsilon=\psi-\varphi$ in the metric $B(r)=1-\frac
{2M}{r}+\gamma r$ under the approximation $\frac{M}{R}<<1$. To this end, we
first determine the value of a small nonzero $\varphi$ lying on the null
geodesic at $r=\infty$ using the orbit Eq.(4). For small $\varphi$, we take
$\sin\varphi\simeq\varphi$, $\cos\varphi\simeq1$ and neglect terms of
$\varphi^{2}$ and higher. Then, at $r=\infty$, the orbit Eq.(4) yields%
\begin{align}
0  &  =-\frac{\gamma}{2}+\frac{\varphi}{R}+\frac{M}{4R^{2}}[8+3R^{2}\gamma
^{2}-3R\gamma(\pi-2\varphi)-6R\gamma\varphi]\nonumber\\
&  -\frac{3M^{2}}{32R^{3}}[128R\gamma+24R^{3}\gamma^{3}-10(2+3R^{2}\gamma
^{2})(\pi-2\varphi)\nonumber\\
&  -14\varphi-16\pi R\gamma\varphi-30R^{2}\gamma^{2}\varphi+3\pi^{2}%
R^{2}\gamma^{2}\varphi].
\end{align}
Solving for $\varphi$, we obtain
\begin{equation}
\varphi=\frac{8MR(8-3\pi R\gamma+3R^{2}\gamma^{2})-16R^{3}\gamma+6M^{2}%
\{5\pi(2+3R^{2}\gamma^{2})-4R\gamma(16+3R^{2}\gamma^{2})\}}{M^{2}(78-48\pi
R\gamma+90R^{2}\gamma^{2}+9\pi^{2}R^{2}\gamma^{2})-32R^{2}}.
\end{equation}
Calculating for small $\psi$, we find%
\begin{align}
\psi &  \simeq\tan\psi=\frac{rB^{1/2}}{r^{2}\frac{du}{d\varphi}}\nonumber\\
&  =\left(  \frac{du}{d\varphi}\right)  ^{-1}\sqrt{\frac{1}{r^{2}}-\frac
{2M}{r^{3}}+\frac{\gamma}{r}},
\end{align}
which goes to $0$ as $r\rightarrow\infty$ since $\frac{du}{d\varphi}\neq0$ at
the value of $\varphi$ derived in Eq.(29). Thus, $\psi=0$ and the one way
deflection then is $\epsilon=0-\varphi$, which easily expands to
\begin{equation}
\epsilon\simeq\frac{2M}{R}\left[  1+\frac{15\pi M}{16R}\right]  -\gamma\left[
\frac{R}{2}+\frac{3\pi M}{4}+\frac{423M^{2}}{32R}\right]  .
\end{equation}

We have checked that this result exactly coincides with that obtained by the
perturbative Bodenner-Will perturbative method (2003). We find that all terms
in the second square bracket are positive, meaning that the effect of
$\gamma>0$ is to diminish (and $\gamma<0$ is to enhance) the Schwarzschild
bending even up to second order in $M$. We also find that Eq.(31) nicely
reproduces the one way deflection $\left(  \epsilon=\frac{2M}{R}-\frac{\gamma
R}{2}\right)  $ obtained by Edery and Paranjape (1998) using Weinberg's
method. As mentioned, the same result (31) follows also from the unaltered
Rindler-Ishak prescription (2007) as well. We shall now discuss some physical
implications of Eq.(31).

\textbf{VI. Physical implications}

First note that in the halo we have obtained a new coupling term $\frac
{3\pi\gamma M}{2}$ between $M$ and $\gamma$, which is independent of the
trajectory parameter $R$. Next, the term $-\frac{\gamma R}{2}$ shows repulsion
for $\gamma>0$, which is consistent with time delay investigations (see e.g.,
Edery \& Paranjape 1998) and attraction if we choose $\gamma<0$. We emphasize
that we are not concluding anything about the correct sign of $\gamma$, which
must be decided by independent observations. When $M=0$ and $\gamma>0$, we
obtain a negative (repulsive) bending of light or $\epsilon=-\frac{\gamma
R}{2}$, which coincides with the conclusion by Walker (1994).

On the other hand, in the galactic halo region, where $R>R_{\text{E}}$ (the
Einstein radius) and $\frac{M}{R}<<1$, one would like to obtain a positive
(attractive) light bending there. This is possible only if one assumes the
condition%
\begin{equation}
\epsilon>0\Rightarrow\frac{2M}{R}>\frac{\gamma R}{2}%
\end{equation}
to hold. Accurately observed lensing data by galactic clusters are now
available. We then find from Table I that the observed values of $M$ and
$R_{\text{E}}$ do indeed respect the inequality (32). Clearly, even if pure
$\gamma>0$ leads to repulsion, in the competition between this repulsion and
Schwarzschild attraction, the latter might win leading to the impression of an
\textit{overall attractive} bending. This can happen in the lensing by
galactic clusters, as described in the table below.

Lens data for $M$, $R_{\text{E}}$ and references are taken from Ishak et
al.\textit{ }(2008), and converted here to length units using $M_{\odot
}=1.475\times10^{5}$cm, $1$ kpc $=$ $3.0856\times10^{21}$cm. The references
are as follows.1: Abell 2744 (Smail et al. 1991, Allen 1998), 2: Abell 1689
(Allen 1998, Limousin 2007), 3: SDSS J1004+4112 (Sharon 2006),4: 3C 295 (Wold
et al. 2002), 5: Abell 2219L (Smail et al. 1995a; Allen 1998), 6: AC 114
(Smail et al. 1995b; Allen 1998). We shall take the rotation curve fit value
$\gamma=3.06\times$ $10^{-30}$ cm$^{-1}$ purely for illustrative purposes and
$\Lambda=1.29\times10^{-56}$cm$^{-2}$ in both the tables below:

\begin{center}
\textbf{Table I }%

\begin{tabular}
[c]{lllll}%
Cluster & $M$ (cm) & $R_{\text{E}}$(cm) & $\frac{2M}{R_{\text{E}}}$ &
$\frac{\gamma R_{\text{E}}}{2}$\\
Abell 2744 & $2.91\times10^{18}$ & $2.97\times10^{23}$ & $1.95\times10^{-5}$ &
$0.45\times10^{-6}$\\
Abell 1689 & $1.38\times10^{18}$ & $4.26\times10^{23}$ & $0.65\times10^{-5}$ &
$0.65\times10^{-6}$\\
SDSS J1004+4112 & $6.28\times10^{18}$ & $3.39\times10^{23}$ & $3.70\times
10^{-5}$ & $0.51\times10^{-6}$\\
3C 295 & $10.5\times10^{18}$ & $3.94\times10^{23}$ & $5.32\times10^{-5}$ &
$0.60\times10^{-6}$\\
Abell 2219L & $4.75\times10^{18}$ & $2.66\times10^{23}$ & $3.57\times10^{-5}$
& $0.40\times10^{-6}$\\
AC 114 & $1.36\times10^{18}$ & $1.68\times10^{23}$ & $1.61\times10^{-5}$ &
$0.25\times10^{-6}$%
\end{tabular}

\end{center}

\ It is evident from the above table that the term $\frac{\gamma R_{\text{E}}%
}{2}$ is smaller than the Schwarzschild term $\frac{2M}{R_{\text{E}}}$, so
that the overall bending is always attractive for $\gamma>0$. One might want
to have an idea of the radius $R$ where the leading order Schwarzschild and
$\gamma-$ bendings balance each other. The value of $R_{\text{b}}$ may be
taken to demarcate the boundary of the halo dark matter surrounding each
individual cluster. This happens at
\begin{equation}
R_{\text{b}}=2\sqrt{\frac{M}{\gamma}}\text{ cm.}%
\end{equation}
The deflection $\epsilon$ below $R<R_{\text{b}}$ is always attractive, as
should be the case. \ Table II shows that the halo boundary $R_{\text{b}}$ can
be several times larger than $R_{\text{E}}$. However, the values of
$R_{\text{b}}$ tabulated here rely crucially on the value of $\gamma$ and if
its value is lowered by one order of magintude than considered here,
$R_{\text{b}}$ will increase by that order. Conversely, if one particular halo
boundary is observationally determined, then it would provide us with a
determination of $\gamma$. One could then examine if that new value of
$\gamma$ explains $R_{\text{b}}$ of other clusters. If it does, then it would
support Weyl theory. Observations seem as yet far too inconclusive about the
sizes of the halo.

\begin{center}
\textbf{Table II }%

\begin{tabular}
[c]{llll}%
Cluster & $M$ (cm) & $R_{\text{E}}$(cm) & $R_{\text{b}}$ (cm)\\
Abell 2744 & $2.91\times10^{18}$ & $2.97\times10^{23}$ & $19.5\times10^{23}$\\
Abell 1689 & $1.38\times10^{18}$ & $4.26\times10^{23}$ & $13.4\times10^{23}$\\
SDSS J1004+4112 & $6.28\times10^{18}$ & $3.39\times10^{23}$ & $28.6\times
10^{23}$\\
3C 295 & $10.5\times10^{18}$ & $3.94\times10^{23}$ & $37.0\times10^{23}$\\
Abell 2219L & $4.75\times10^{18}$ & $2.66\times10^{23}$ & $24.9\times10^{23}%
$\\
AC 114 & $1.36\times10^{18}$ & $1.68\times10^{23}$ & $13.3\times10^{23}$%
\end{tabular}

\end{center}

Though our interest so far has only been in the galactic clusters, one might
still want to compare the magnitudes of the $\gamma-$ related effects with the
Schwarzschild ones in the solar system although the region around the Sun
contains galactic matter. For a light ray grazing the Sun, we have the
following numerical values:%
\begin{equation}
M_{\odot}=1.475\times10^{5}\text{ cm, }R_{\odot}=6.96\times10^{10}\text{ cm,
}\gamma=3.06\times10^{-30}\text{cm}^{-1}%
\end{equation}
so that
\begin{equation}
\frac{2M_{\odot}}{R_{\odot}}=4.24\times10^{-6}\text{, \ }\frac{30\pi M_{\odot
}^{2}}{16R_{\odot}^{2}}=2.65\times10^{-11}\text{,}%
\end{equation}%
\begin{equation}
\frac{\gamma R_{\odot}}{2}=1.06\times10^{-19}\text{, \ }\frac{3\pi\gamma
M_{\odot}}{4}=1.06\times10^{-24}\text{.}%
\end{equation}
We find that the $\gamma-$ correction terms are considerably small compared to
$\frac{2M_{\odot}}{R_{\odot}}$, therefore the effect of $\gamma$ would be
negligible near the Sun. However, as illustrated in Table I, the effect of
$\gamma$ near any galactic cluster scale is not as negligible. The fact that
$\gamma$ is meaningful only on such large scales has been conjectured in the
literature, but here we have found its support from a completely different
viewpoint, viz., from the Rindler-Ishak bending.

\textbf{VII. Summary and results}

We calculated light deflection in the vacuole model up to third order in $M$
and confirmed that the extension of Rindler-Ishak method to the
Einstein-Strauss vacuole, as originally developed by Ishak et al. (2008),
reproduces the Schwarzschild $M-$ dependent bending terms as well as the
$\Lambda$ $-$ dependent terms, see Eq.(21). In particular, we have found a
local coupling term $-\frac{2M\Lambda r_{0}}{3}$ similar to that by Sereno. We
have also found a more interesting coupling term $-\frac{5\pi M^{2}\Lambda}%
{8}$ including other new terms, the most notable one being $-\frac{\Lambda
Rr_{b}}{3}$. It would be of interest to discuss the recessional impact too
(Ishak \& Rindler, 2010), but it requires a separate and detailed investigation.

The idea of a cut-off transition region between the halo boundary and the
exterior dS cosmology was conjectured, but not implemented, by Edery \&
Paranjape (1998) over a decade ago. The SdS vacuole model by Ishak et al.
(2008) is philosophically the same in idea but different in content. It
envisages a transition radius $r_{b}$ between the SdS vacuole boundary and the
exterior FRW cosmology implementing the Einstein-Strauss suggestion. The
vacuole surrounding the lens should be deviod of matter, and therefore the
model particularly applies to galactic clusters rathen than local objects like
the Sun, which is surrounded by galactic matter.

We have argued that the vacuole method is exclusive to cases where the
cosmological constant $\Lambda$ disappears from the second order differential
path equation. To exemplify it, we applied the vacuole model in the
calculation of the $\gamma-$ dependent effects in Weyl gravity. We note that
the parameter $\gamma$ does not disappear from the path equation, and thus the
vacuole method does not yield the known Weyl term $-\frac{\gamma R}{2}$. To
this end, we point out that the earlier Rindler-Ishak (2007) prescription in
their non-vacuole method did nicely yield the otherwise known Weyl term
(Bhattacharya et al. 2010). In the present paper, we showed that an
alternative prescription on the azimuthal angle lying on the null orbit also
reveal the influence of the Schwarzschild ($M$) and conformal sector ($\gamma
$) on light deflection [See Eq. (31)]: It reproduced the correct Schwarzschild
bending terms due to $M>0$ as well as those due to the conformal Weyl
parameter $\gamma$. In particular, the known term $-\frac{\gamma R}{2}$
followed exactly. Also we have found a new local coupling term $\frac
{3\pi\gamma M}{2}$ between $M$ and $\gamma$, which is independent of the
trajectory parameter $R$. We chose (not mandatorily) the value obtained by
Mannheim (2006) from the fit of the galactic flat rotation curve data and
applied it to the accurately observed data on several galactic clusters taken
from Ishak et al. (2008). We have shown in Table I that, for $R_{\text{E}}\leq
R<R_{\text{b}}$, the light bending is attractive since\ $\ \epsilon$ $\left(
=\frac{2M}{R}-\frac{\gamma R}{2}\right)  $ is always positive \textit{masking}
the purely negative Weyl $\gamma-$ term, while Table II gives possible sizes
$R_{\text{b}}$ of the halo if the chosen value of $\gamma$ is relied upon.
Although galactic halo can be modelled in many ways [see, for instance, the
brane world model, Nandi et al (2009)], the interpretations of Weyl gravity in
this regime seem as yet conclusive, to our knowledge.

\textbf{Acknowledgment}

The authors are deeply indebted to Guzel N. Kutdusova for her encouragement
and assistance.

\textbf{References}

Allen S., 1998, Mon. Not. Roy. Astron. Soc. \textbf{296}, 392

Bhadra A., Sarkar K and Nandi K.K., 2007, Phys. Rev. D \textbf{75},123004

Bhattacharya A. et al., \ 2010, JCAP 09:004

Bodenner J. and Will C.M., 2003, Am. J. Phys. \textbf{71}, 770

Carroll S., 2001, Living Reviews in Relativity, \textbf{4}, 1

Darmois G.,1927, \textit{M\'{e}morial de Sciences Math\'{e}matiques, Fascicule
XXV}, "Les equations de la gravitation Einsteinienne", Chapitre V

Edery A. and Paranjape M.B., 1998, Phys. Rev. D, \textbf{58}, 024011

Einstein A. and Strauss E., 1945, Rev. Mod. Phys. \textbf{17}, 120. Erratum:
1946, ibid \textbf{18}, 148

Ishak M. et al., 2008, Mon. Not. Roy. Astron. Soc. \textbf{388}, 1279

Ishak M., Found. Phys., 2007, \textbf{37}, 1470

Ishak M. and \ Rindler W., 2010, arXiv:1006.0014, and references therein, to
appear in GRG.

Ishak M., \ Rindler W. and Dossett J., 2010, Mon. Not. Roy. Astron. Soc.
\textbf{403}, 2152

Israel W., 1966, Nuovo Cim. B \textbf{44}, 1. Erratum: 1967, ibid \textbf{48}, 463

Lanczos C., 1924, Ann Phys. (Leipzig), \textbf{74}, 518

Limousin M. et al., 2007, Astrophys. J. \textbf{668}, 643

Mannheim P.D. and Kazanas D., 1989, Astrophys. J. \textbf{342}, 635

Mannheim P.D., 1997, Astrophys. J. \textbf{479}, 659

Mannheim P.D., 2006, Prog. Part. Nucl. Phys. \textbf{56}, 340

Nandi K.K. \textit{et al}., 2009, Mon. Not. R. Astron. Soc. \textbf{399}, 2079

Page L. et al., 2003, Astrophys. J. Suppl.\textbf{148}, 2333

Peebles J. and Ratra B., 2003, Rev. Mod. Phys.\textbf{ 75}, 559

Perlmutter S. et al., 1999, Astrophys. J. \textbf{517}, 565

Pireaux S., 2004, Class. Quant. Grav. \textbf{21}, 4317

Riess A., et al., 1998, Astron. J. \textbf{116}, 1009

Rindler W., 1969, Astrophys. J. \textbf{157}, L147

Rindler W. and Ishak M., 2007, Phys. Rev. D \textbf{76}, 043006

Sen N.R., 1924, Ann. Phys. (Leipzig), \textbf{73}, 365

Sereno M, 2008, Phys. Rev. D \textbf{77}, 043004

Sereno M., 2009, Phys. Rev. Lett. \textbf{102}, 021301

Sharon K. et al., 2005, Astrophys. J. \textbf{629}, L73

Smail I. et al., 1991, Mon. Not. Roy. Astron. Soc. \textbf{252}, 19

Smail I. et al., 1995, Mon. Not. Roy. Astron. Soc. \textbf{277}, 1

Smail I. et al., 1995, Astrophys. J. \textbf{440}, 501

Spergel D., et al., 2007, Astrophys. J. Suppl. \textbf{170}, 377

Walker M.A. , Astrophys. J. \textbf{430}, 463 (1994).

Weinberg S., 1972, \textit{Gravitation \& Cosmology }(John Wiley \&Sons, New York)

Will C.M., 2001, Living Rev. Relativ. \textbf{4}, 2001-2004

Wold M. et al., 2002, Mon. Not. Roy. Astron. Soc. \textbf{335}, 1017

\bigskip

\begin{center}
\textbf{Appendix}
\end{center}

The integration of the first order equation reads%
\begin{equation}
\varphi^{\text{Sereno}}=\pm\int\frac{dr}{r^{2}}\left[  \frac{1}{b^{2}}%
+\frac{\Lambda}{3}-\frac{1}{r^{2}}+\frac{2M}{r^{3}}\right]  ^{-1/2}.=\pm\int
f(M,\Lambda,b,r)dr \tag{A1}%
\end{equation}
It can't be integrated in a closed form. So expanding the integrand $f$ in
first power of $M$, we have%
\begin{equation}
f=\frac{1}{r^{2}}\left(  \frac{1}{b^{2}}-\frac{1}{r^{2}}-\frac{\Lambda}%
{3}\right)  ^{-1/2}-\frac{M}{r^{5}}\left(  \frac{1}{b^{2}}-\frac{1}{r^{2}%
}-\frac{\Lambda}{3}\right)  ^{-3/2}\equiv f_{1}+f_{2}\text{ \ (say).} \tag{A2}%
\end{equation}
Then, to first power of $\Lambda$,%
\begin{align}
I_{1}  &  =\int f_{1}dr\nonumber\\
&  =\frac{\sqrt{b^{2}(3+r^{2}\Lambda)-3r}\left[  \ln r-\ln2-\ln\left\{
b\sqrt{3}+\sqrt{b^{2}(3+r^{2}\Lambda)-3r}\right\}  \right]  }{\sqrt{3}%
r\sqrt{1-\frac{b^{2}}{r^{2}}-\frac{\Lambda b^{2}}{3}}}\nonumber\\
&  \simeq-\frac{b}{r}-\frac{b^{3}}{6r^{3}}-\frac{3b^{5}}{40r^{5}}%
-\frac{\Lambda b^{3}}{6r}-\frac{\Lambda b^{5}}{12r^{3}}+\text{imaginary
terms.} \tag{A3}%
\end{align}%
\begin{align}
I_{2}  &  =\int f_{2}dr\nonumber\\
&  =-\frac{M[b^{2}(3+2r^{2}\Lambda)-6r^{2}]}{3br^{2}\sqrt{1-\frac{b^{2}}%
{r^{2}}-\frac{\Lambda b^{2}}{3}}}\nonumber\\
&  \simeq\frac{2M}{b}-\frac{M\Lambda b}{3}+\frac{Mb^{3}}{4r^{4}}+\frac{Mb^{5}%
}{4r^{6}}+\frac{M\Lambda b^{5}}{8r^{4}}. \tag{A4}%
\end{align}
Collecting real terms, we get%
\begin{align}
\varphi^{\text{Sereno}}  &  =\frac{2M}{b}-\frac{b}{r}-\frac{M\Lambda b}%
{3}-\frac{b^{3}}{6r^{3}}+\frac{Mb^{3}}{4r^{4}}-\frac{3b^{5}}{40r^{5}}%
-\frac{\Lambda b^{3}}{6r}\nonumber\\
&  -\frac{M\Lambda^{2}b^{3}}{36}-\frac{\Lambda b^{5}}{12r^{3}}+\frac{Mb^{5}%
}{4r^{6}}+\frac{M\Lambda b^{5}}{8r^{4}}, \tag{A5}%
\end{align}
which seem to yield that the local coupling term is $-\frac{2M\Lambda b}{3}$.

\bigskip

\begin{center}

\bigskip

\bigskip
\end{center}

\bigskip

\begin{center}
\bigskip

\end{center}

\end{document}